\documentclass[apjl]{emulateapj}
\begin{document}
\title{Probing the Warm Dark Matter with 
High-$z$ Quasar Luminosity Function }
\author{Hyunmi Song and Jounghun Lee}
\affil{Department of Physics and Astronomy, FPRD, Seoul National University, 
Seoul 151-747, Korea} 
\email{yesuane@gmail.com}
\email{jounghun@astro.snu.ac.kr}
\begin{abstract}
In a warm dark matter (WDM) cosmology, the first objects to form at $z\ge 20$ 
are one dimensional filaments with mean length on the order of the WDM 
free-streaming scale. Gao and Theuns recently claimed by using  
high-resolution hydrodynamic simulations that the eventual collapse of 
these WDM filaments along their longest axes may seed the supermassive black 
holes that power high-$z$ quasars. In this picture, it is supposed that the 
high-$z$ quasar luminosity function should reflect how abundant the WDM 
filaments are in the early universe. We derive analytically the mass function 
of early-universe filaments with the help of the Zel'dovich approximation. 
Then, we determine the rate of its decrease in the mass section corresponding 
to the free streaming scale of a WDM particle of mass $m_{\nu}$. Adjusting the 
value of $m_{\nu}$, we fit the slope of the analytic model to that 
of the high-$z$ quasar luminosity function measured from the Sloan 
Digital Sky Survey DR3. A new WDM constraint from this feasibility study 
is found to be consistent with the lightest super-symmetric partner.
\end{abstract}
%%%%%%%%%%%%%%%%%%%%%%%%%%%%%%%%%%%%%%%%%%%%%%%%%%%%%%%%%%%%%%%%%%%%%%%%%%%

\section{INTRODUCTION}

The large-scale features of the observed universe are strikingly consistent 
with the theoretical predictions based on the cold dark matter model. 
The combined analyses of the recent data from the observations of cosmic 
microwave background (CMB), galaxy power spectrum and  Type Ia supernovae 
\citep[e.g.,][and references therein]{wmap5} have been capable of measuring 
the key cosmological parameters that characterize the CDM model with 
surprisingly high precision. This has opened an era of precision cosmology, 
echoing the triumph of the CDM model. 

Nevertheless, the status of the CDM model as the standard paradigm has been 
shaking currently in both observational and theoretical perspectives.
Observations have reported several mismatches between the predictions of the 
CDM model and the real phenomena on galactic and subgalactic scales. For 
instance, the abundance of galactic satellites, the slope of the inner core 
of the dark halo density profiles, and the degree of void emptiness have 
exhibited apparent conflicts between theory and observation 
\citep{kly-etal99,moo-etal99,pee01}. 
Very recently, \citet{dis-etal08} measured the cross-correlations between 
different galaxy's observables and showed that only a single parameter 
suffices to explain the complex structures of galaxies. This phenomenon is 
inconsistent with the CDM picture where the galaxy formation is driven by 
hierarchical merging process. Although it is still inconclusive whether the 
reported mismatches really indicate the failure of the CDM model or they are 
just by-products of the selection effects and/or limitations of the 
observational techniques, it has certainly led to an emergence of 
alternative models.

A theoretical challenge comes from the fact that the extension of the standard 
particle physics favors the warm dark  matter (WDM) rather than CDM. 
The WDM differs from CDM in a respect that the WDM particles have 
non-negligible free streaming scale, $l_{\nu}$. 
On scales larger than $l_{\nu}$, WDM behave just like CDM. On the other 
hand, on scales less than $l_{\nu}$, WDM do not cluster but free stream 
due to their large velocity dispersions unlike CDM which cluster on all 
scales. It is this property that enables the WDM model to overcome the 
observational mismatches faced by the CDM model. 
The list of realistic WDM candidates that have so far proposed includes 
gravitinos \citep[][and references therein]{ber-etal05}, axions \citep{axion}, 
and sterile neutrinos \citep{DW94}, each of which has distinct characteristic 
mass range. By constraining the WDM particle mass, it may be possible to rule 
out a certain WDM candidate. In fact there have been already such attempts: 
For instance, \citet{sel-etal06} constrained the particle mass of WDM by 
using the Ly$\alpha$ forest power spectrum and suggested that the sterile 
neutrinos be ruled out \citep[see also][]{vie-etal06} 

In a WDM universe, the objects that will first condense out in the initial 
density inhomogeneities are not zero dimensional clumps but one dimensional 
filaments. \citet{GT07} recently studied the formation and properties of 
WDM filaments with high-resolution hydrodynamic simulations, and proposed 
that the early-universe filaments would seed the supermassive black holes 
that power the quasars at $z\ge 6$ through their eventual collapse 
along the longest axes.  If this scenario is true, then, the following two 
issues should be related to each other: How luminous and abundant the first 
generation high-$z$ quasars are; and how massive and abundant the 
early-universe WDM filaments are. In statistical terms, the luminosity 
function of the high-$z$ quasars should be related to the mass function of 
the early-universe filaments. Given that the mass function of the 
early-universe filaments depends on the particle mass of WDM (see \S 2), 
it implies the possibility of using the high-$z$ quasar luminosity 
function as a new WDM constraint. 

The idealistic way to study the mass function of early-universe filaments 
and its dependence on the WDM particle mass is to use the hydrodynamics 
simulations. However,  given the computational costs as well as the current 
resolution-limit of hydrodynamic simulations,  it is adequate and quite 
necessary to consider an analytic approach as the first guideline.
Here, we evaluate analytically the abundance of early-universe filaments and 
constrain the WDM particle mass by fitting its slope in the WDM particle 
free streaming scale to that of the high-$z$ quasar luminosity function 
\citep{fan-etal01} determined from the Sloan Digital Sky Survey Data 
Release 3 (SDSS) \citep{sdssdr3}. For the key cosmological parameters other 
than the WDM particle mass, we assume a WMAP5 cosmology \citep{wmap5} 
throughout this paper.

The plan of this paper is as follows.
In \S 2, an analytic model for the abundance of early universe filaments is  
presented. In \S 3, a constraint on the particle mass of WDM is provided by 
fitting the analytic model to the observational data from SDSS.
In \S 4, the achievements as well as the caveats of our work are discussed. 

\section{THE ABUNDANCE OF EARLY UNIVERSE FILAMENTS}

The Zel'dovich approximation predicts generically and in the most simple way  
the formation of early-universe filaments \citep{zel70}. 
According to this model, the mass density $\rho$ of a given region can be 
expressed in terms of the Lagrangian quantities as
\begin{equation}
\rho = \frac{\bar{\rho}}{[1-D(z)\lambda_1][1-D(z)\lambda_2][1-D(z)\lambda_3]}, 
\label{eqn:rho}
\end{equation}
where $\bar{\rho}$ is the mean background density, $\lambda_{1},\lambda_{2},
\lambda_{3}$ (in a decreasing order) are the three eigenvalues of the 
linear deformation tensor, and $D(z)$ is a linear growth factor. Equation 
(\ref{eqn:rho}) predicts that $\rho$ will diverge at $\lambda_{1}=1/D(z)$. 
That is, the first collapse will occur when the largest eigenvalue reaches 
a threshold value, $\lambda_{c}(z)\equiv 1/D(z)$. The sign of the other two 
eigenvalues, $\lambda_{2}$ and $\lambda_{3}$, at the moment of the first  
collapse will determine the dimension of a collapsed object. 
A one-dimensional filament will form if $\lambda_{2}>0$ and $\lambda_{3}<0$, 
while the formation of a two-dimensional sheet will occur for the case 
$\lambda_{2}<0$ and $\lambda_{3}<0$. Since we are interested in the 
formation of ``filaments'' that are shown to be most abundant in the early 
universe filled by WDM \citep{GT07}, we consider only the filament condition, 
$\lambda_{1}=1/D(z)$, $\lambda_{2}>0$ and $\lambda_{3}<0$.

The fractional volume $F(M,z)$ occupied by those Lagrangian regions which 
will condense out early-universe filaments at $z$ on mass scale $M$ can be 
expressed as
\begin{equation}
\label{eqn:fm}
F(M,z) = p[\lambda_{1}\ge\lambda_{c}(z),\lambda_{2}\ge 0,
\lambda_{3}\le 0:\sigma_{M}],
\end{equation}
where $\lambda_{c}(z)=1/D(z)$ and $\sigma_{M}$ is the rms density fluctuation 
related to the dimensionless linear WDM power spectrum $\Delta^{2}_{\nu}(k)$ 
as 
\begin{equation}
\sigma^{2}_{R} \equiv \int_{-\infty}^{\infty}\Delta^{2}_{\nu}(k)
W^{2}(k,M)d\ln k, 
\end{equation}
where $W(k,M)$ represents a sharp k-space filter on mass scale $M$. 
We adopt the approximation formula of\citet{BO01}, $\Delta^{2}_{\nu}(k)=
\Delta^{2}(k) \big\vert T_{k}^{\nu} \big\vert^{2}$
where $\Delta^{2}(k)$ is the CDM linear power spectrum. The 
functional form of the WDM transfer function, $T_{k}^{\nu}$ is given as  
\begin{equation}
T_{k}^{\nu}=\left[1+(\alpha k)^{2\gamma}\right]^{-5/\gamma}
\end{equation}
where
\begin{equation}
\alpha=0.048(\Omega_{\nu}/0.4)^{0.15}(h/0.65)^{1.3}
            ({\rm keV}/m_{\nu})^{1.15}(1.5/g_{\nu})^{0.29}, 
\end{equation}
$\gamma=1.2$, $m_{\nu}$ is the WDM particle mass and $g_{\nu}$ is the number 
of degrees of freedom that equals $3/4$ for fermions. 

The fractional volume in equation (\ref{eqn:fm}) can be obtained by 
integrating the joint probability density distribution of the three 
eigenvalues that was first derived by \citet{dor70} as
\begin{eqnarray}
p(\lambda_1,\lambda_2,\lambda_3;\sigma_{M})&=&
\frac{3375}{8\sqrt{5}\pi\sigma^6_{M}}
\exp\bigg{(}-\frac{3I_{1}^2}{\sigma^2_{M}}
+\frac{15I_{2}}{2\sigma^2_{M}}\bigg{)}\\ \nonumber
&\times&(\lambda_{1}-\lambda_{2})(\lambda_{2}-\lambda_{3})
(\lambda_{1}-\lambda_{3}), 
\label{eqn:dor}
\end{eqnarray}
where $I_{1}=\lambda_{1}+\lambda_{2}+\lambda_{3}$ and 
$I_{2}=\lambda_{1}\lambda_{2}+\lambda_{2}\lambda_{3}+\lambda_{3}\lambda_{1}$.
In the spirit of the Press-Schechter theory \citep[][hereafter PS]{PS74}, 
we evaluate the number density of early-universe filaments collapsed at 
mass scale $M$ as
\begin{equation}
\label{eqn:ps}
n_{\rm F}(M,z) = \frac{\bar{\rho}}{M}\left\vert\frac{d}{dM}F(M,z)\right\vert, 
\label{eqn:nf}
\end{equation} 
where the redshift for the formation of the early-universe filaments is 
set at $z=23$, in accordance with \citet{GT07}. In the original PS theory 
which dealt with the mass function of the clumps, equation (\ref{eqn:nf}) 
had to be multiplied by an additional normalization constant to account for 
the occurrence of the cloud-in-clouds \citep{bon-etal91}. 
But, for our case the cloud-in-clouds will not occur since the WDM particles 
just free-stream in the early-universe filaments forming no substructures. 
Henceforth, we do not have to account for the cloud-in-cloud problem in 
deriving the mass function of early universe filaments. 

\begin{figure} 
\plotone{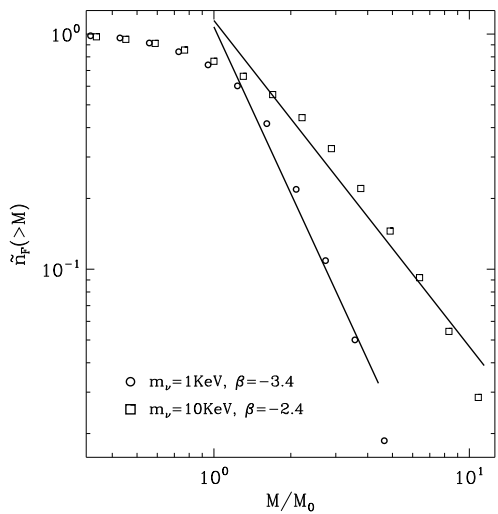}
\caption{The rescaled cumulative mass function of early-universe filaments 
for the two different cases of WDM particle mass ($m_{\nu}=1$ and $10$keV as 
open circles and open squares, respectively). The value $M_{0}$ corresponds 
to the free-streaming mass scale of a given WDM particle. The solid lines 
represent the power-law fitting in the mass range where 
$0.1\le n_{F}(>M)/n_{F}(>M_{0})\le 1$ for the two cases.}
\label{fig:nf}
\end{figure}

The cumulative mass function $n_{\rm F}(>M)$ of early universe filaments 
can be obtained by integrating equation (\ref{eqn:nf}) as 
$n_{\rm F}(>M)\equiv \int_{M}^{\infty}n_{\rm F}(M)dM$. Since our focus 
is not on the amplitude of the mass function but mainly on its slope, 
we rescale $n_{\rm F}(>M)$ to have unity as $M$ goes zero.
Figure \ref{fig:nf} plots the rescaled cumulative mass function 
$\tilde{n}_{\rm F}(>M)$ at $z=23$ versus $M/M_{0}$ for the case of 
$m_{\nu}=1$keV (open circles) and $m_{\nu}=10$keV (open squares). 
Here $M_{0}$ corresponds to the free streaming mass scale of a given WDM 
particle.  
%%%%%%%%%%%%%%%%%%%%%%%%%%%%%%%%%%%%%%%%%%%%%%%%%%%%%%%%%%%%%%%%%%%%%%%%%%%%
%%%%%%%%%%%%%%%%%%%%%%%%%%%%%%%%%%%%%%%%%%%%%%%%%%%%%%%%%%%%%%%%%%%%%%%%%%
As can be seen, $n_{\rm F}(>M)$ begins to decrease as $M$ becomes larger than 
$M_{0}$.  The rate of its decrease at $M > M_{0}$ depends sensitively on 
the value of the WDM particle mass $m_{\nu}$: It decreases more rapidly with 
$M$ for the case of smaller value of $m_{\nu}$. It is because the less 
massive WDM particles have smaller free-stream scales. 
To quantify the rate of the decrease of $n_{\rm F}(>M)$ at $M > M_{0}$, 
we fit $n_{F}(>M)$ to a power-law $M^{\beta+1}$ 
(i.e., $n_{F}(M)\propto M^{\beta}$) 
in the mass range where $0.1\le n_{\rm F}(>M)/n_{\rm F}(>M_{0})\le 1$.
In the higher mass section where $n_{\rm F}(> F)$ drops exponentially 
with $M$, we expect basically no early universe filaments. 
As demonstrated by \citet{GT07}, only those early-universe filaments with 
mass comparable in order of magnitude to the WDM free-streaming scale can 
form which are in turn related to the first-generation supermassive black 
holes that power the observed high-$z$ luminous quasars. The higher mass 
section where $n_{\rm F}(>M)/n_{\rm F}(>M_{0})< 0.1$ correspond to the scales 
order-of-magnitude larger than the free streaming scale of a given 
WDM particles.  The solid lines in Fig.~\ref{fig:nf} represent the 
power-law fitting of the cumulative mass function of early universe 
filaments. As can be seen, the slope $\beta$ indeed depends sensitively 
on the value of $m_{\nu}$.

\section{RELATION to HIGH-$z$ QUASAR LUMINOSITY FUNCTION}

The discovery of a supermassive black hole with mass $3\times 10^{9}M_{\odot}$ 
in quasar SDSS J1148+5251 at $z=6.41$ \citep{wil-etal03} has raised a 
crucial question of what seeded such a supermassive black hole at that 
early epoch \citep[][and references therein]{pod-etal03}. 
As introduced in \S 1, 
\citet{GT07} have studied the formation of early-universe filaments using 
high-resolution hydrodynamic simulations for a WDM cosmology and suggested 
that the ultimate collapse of the WDM filaments along the longest axes should 
induce vigorous collisions between gas clouds and stars due to their high 
densities, which would in turn seed the formation of supermassive black holes 
that power quasars at $z\ge 6$. 
In subsequent evolution, the black holes would grow active by accreting 
baryonic gases and dark matter particles that constituted their parent 
filaments. In this picture, the number density of active black holes at 
high redshifts would be related to that of early universe filaments.

%%%%%%%%%%%%%%%%%%%%%%%%%%%%%%%%%%%%%%%%%%%%%%%%%%%%%%%%%%%%%%%%%%%%%%%%%%%%
\begin{figure} 
\plotone{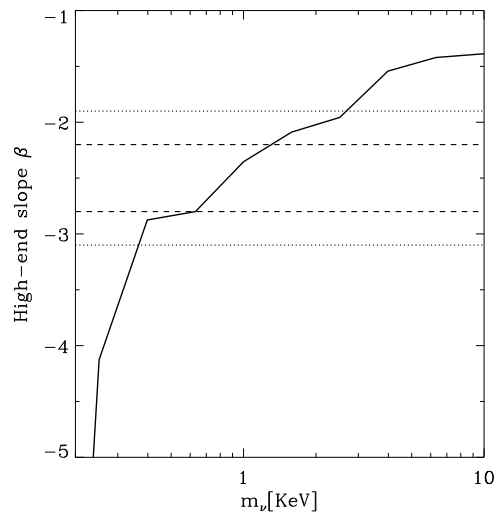}
\caption{The power-law slope of the mass function of early universe 
filaments versus WDM particle mass. The horizontal dashed and dotted lines 
correspond to the $1\sigma$ and $2\sigma$ ranges of the high-end slope of 
high-$z$ quasar luminosity function determined from SDSS DR3 
\citep{fan-etal01},respectively.}
\label{fig:slope}
\end{figure}
%%%%%%%%%%%%%%%%%%%%%%%%%%%%%%%%%%%%%%%%%%%%%%%%%%%%%%%%%%%%%%%%%%%%%%%%%%
The luminosity function $\phi~(L)$ of high-$z$ quasars at $z\ge 4$ was 
determined by \citet{fan-etal01}, who showed that $\phi~(L)$ can be well 
approximated as a power law with high-end slope in the range of 
$-2.5\pm 0.25$. If these high-$z$ quasars are indeed powered by the 
first-generation supermassive black holes formed through the ultimate 
collapse of the WDM filaments, the high-end slope of the high-$z$ quasar 
luminosity function should be consistent with the power-law slope of the 
mass function of early-universe filaments determined in \S 2.  
Varying the value of $m_{\nu}$, we determine the power-law slopes of the 
mass function of early-universe filaments $\beta$, which are shown in Figure 
\ref{fig:slope}. The $1\sigma$ and $2\sigma$ range of the high-end slope of 
the high-$z$ quasar luminosity function $\phi~(L)$ measured from SDSS DR3 
are shown as horizontal dashed and dotted lines, respectively. The comparison 
indicates that the two slopes agree with each other when $m_{\nu}$ 
is a few keV.

\section{DISCUSSION AND CONCLUSION}

Motivated by the recent heuristic work of \citet{GT07}, we have constructed 
an analytic model for the mass function of the early-universe filaments in 
a WDM universe. Then we test the possibility of using it as a new WDM 
constraint. As a feasibility study, we compare the high-end slope of the mass 
function of the early-universe filaments to that of the high-$z$ quasar 
luminosity function determined from SDSS DR3 and find a new WDM constraint, 
$m_{\nu}\sim$ a few keV. This preliminary result looks consistent 
with the super-symmetric picture \citep{ber-etal05}. 
The advantage of the mass function of the early-universe filament as a WDM 
constraint lies in the fact that the model is purely analytical, free of any 
fitting or nuisance parameter. It depends only on the WDM particle mass 
other than the key cosmological parameters. In addition, when the analytic 
model is compared to the observational result, it is not required to account 
for any complicated baryon physics since we consider only the {\it slope} of 
the mass function. 

Yet, there are a couple of simplified assumptions on which our model is based.
First, the redshifts of the quasars should be higher than $z\ge 6$ for a 
more fair comparison. In this work, we just used the redshift range $z\ge 4$, 
assuming that the quasars at $z\ge 4$ correspond to the first generation. 
To be more realistic,  it will be required to determine the luminosity 
function of the quasars at $z\ge 6$, but there are only small number of 
quasars have so far been observed at that high redshift. For example, 
\citet{ves-etal08} determined the supermassive black hole mass function at 
$z\ge 4.7$ using a tens of quasars and found the high-end slope in a 
much larger range of $-1.9\pm 1.7$, which obviously suffers from large 
uncertainty due to the small number statistics. Anyway, we would like to 
mention clearly that the true high-end slope of the high-$z$ quasar 
luminosity function is currently significantly more uncertain than 
the data used here.

Second, the analytic model for the mass function of early universe filaments 
has to be refined. Our model is the simplest approximation based on the 
Zel'dovich approximation and the Press-Schechter theory. The validity of 
our model has to be tested numerically before using it in practice. 
Third, the evolution of the early-universe filaments has not been taken 
into account properly for the evaluation of the mass function. In reality, 
before their eventual collapse at $z\ge 6$, the early-universe filaments may 
undergo fragmentation due to the tidal effect from the surrounding matter. 
To account for the evolutionary effects, it will be anyway required to 
refine the model with the help of the hydrodynamic simulations. Our future 
work is in this direction. 

As a final conclusion, our model and the preliminary result from our 
feasibility study has provided a proof of concept that the high-$z$ 
quasar luminosity function will be in principle a useful WDM probe.

\acknowledgments

We thank an anonymous referee for helpful comments.
This work is financially supported by the Korea Science and Engineering 
Foundation (KOSEF) grant funded by the Korean Government 
(MOST, NO. R01-2007-000-10246-0).


\begin{thebibliography}{}
\bibitem[Battye \& Shellard(1994)]{axion}
Battye, R.~A., \& Shellard, E.~P.~S.\ 1994, \prl, 73, 2954 
\bibitem[Bertone et al.(2005)]{ber-etal05}
Bertone, G., Hooper, D., \& Silk, J.\ 2005, \physrep, 405, 279 
\bibitem[Bode et al.(2001)]{BO01} 
Bode, P., Ostriker, J.~P., \& Turok, N.\ 2001, \apj, 556, 93 
\bibitem[Bond et al.(1991)]{bon-etal91}
Bond, J.~R., Cole, S., Efstathious, G., \& Kaiser, N.\ 1991, \apj, 379, 440
\bibitem[Christlieb et al.(2002)]{chr-etal02} 
Christlieb, N., et al.\ 2002, \nat, 419, 904 
\bibitem[Coles et al.(1993)]{col-etal93} 
Coles, P., Melott, A.~L., \& Shandarin, S.~F.\ 1993, \mnras, 260, 765
\bibitem[Croft et al.(1998)]{cro-etal98} 
Croft, R.~A.~C., Weinberg, D.~H., Katz, N., \& Hernquist, L.\ 1998, 
\apj, 495, 44 
\bibitem[Dalla Bont{\`a} et al.(2009)]{DB-etal09}
Dalla Bont{\`a}, E., Ferrarese, L., Corsini, E.~M., Miralda-Escud{\'e}, J., 
Coccato, L., Sarzi, M., Pizzella, A., \& Beifiori, A.\ 2009, \apj, 690, 537 
\bibitem[Disney et al.(2008)]{dis-etal08}
Disney, M.~J., Romano, J.~D., Garcia-Appadoo, D.~A., 
West, A.~A., Dalcanton, J.~J., \& Cortese, L.\ 2008, \nat, 455, 1082 
\bibitem[Dodelson \& Widrow(1994)]{DW94} 
Dodelson, S., \& Widrow, L.~M.\ 1994, \prl, 72, 17 
\bibitem[Doroshkevich(1970)]{dor70} 
Doroshkevich, A.~G.\ 1970, Astrofizika, 6, 581 
\bibitem[Dunkley et al.(2009)]{wmap5}
Dunkley, J., et al.\ 2009, \apjs, 180, 306 
\bibitem[Fan et al.(2001)]{fan-etal01} 
Fan, X., et al.\ 2001, \aj, 121, 54 
\bibitem[Frebel et al.(2005)]{fre-etal05}
Frebel, A., et al.\ 2005, \nat, 434, 871 
\bibitem[Gao \& Theuns(2007)]{GT07} 
Gao, L., \& Theuns, T.\ 2007, Science, 317, 1527 
\bibitem[G{\"o}tz \& Sommer-Larsen(2002)]{got-etal02}
G{\"o}tz, M., \& Sommer-Larsen, J.\ 2002, \apss, 281, 415
\bibitem[Hu et al.(2006)]{hu-etal06}
Hu, J., Shen, Y., Lou, Y.-Q., \& Zhang, S.\ 2006, \mnras, 365, 345  
\bibitem[Klypin et al.(1999)]{kly-etal99}
Klypin, A., Kravtsov, A.~V., Valenzuela, O., \& Prada, F.\ 1999, 
\apj, 522, 82 
\bibitem[Lee \& Shandarin(1998)]{LS98}
Lee, J. \& Shandarin, S.~F.\ 1998, ApJ, 500, 14
\bibitem[Moore et al.(1999)]{moo-etal99}
Moore, B., Ghigna, S., Governato, F., Lake, G., Quinn, T., Stadel, J., 
\& Tozzi, P.\ 1999, \apjl, 524, L19 
\bibitem[Munyaneza \& Biermann(2005)]{MB05}
Munyaneza, F., \& Biermann, P.~L.\ 2005, \aap, 436, 805 
\bibitem[Narayanan et al.(2000)]{nar-etal00} 
Narayanan, V.~K., Spergel, D.~N., Dav{\'e}, R., \& Ma, C.-P.\ 2000, 
\apjl, 543, L103 
\bibitem[Peebles(2001)]{pee01}
Peebles, P.~J.~E.\ 2001, \apj, 557, 495
\bibitem[Podsiadlowski et al.(2003)]{pod-etal03}
Podsiadlowski, Ph., Rappaport, S. \& Hau, Z. 2003, \mnras, 341, 385 
\bibitem[Press \& Schechter(1974)]{PS74} 
Press, W.~H., \& Schechter, P.\ 1974, \apj, 187, 425 
\bibitem[Seljak et al.(2006)]{sel-etal06} 
Seljak, U., Makarov, A., McDonald, P., \& Trac, H.\ 2006, 
\prl, 97, 191303 
\bibitem[Trump et al.(2006)]{sdssdr3}
Trump, J.~R., et al.\ 2006, \apjs, 165, 1 
\bibitem[Vestergarrd et al.(2008)]{ves-etal08}
Vestergaard, M., Fan, X., Tremonti, C.~A., Osmer, P.~S., 
Richards, G.~T.\ 2008, \apjl, 674, L1
\bibitem[Viel et al.(2006)]{vie-etal06}
Viel, M., Lesgourgues, J., Haehnelt, M.~G., Matarrese, S., 
\& Riotto, A.\ 2006, \prl, 97, 071301
\bibitem[Willott et al.(2003)]{wil-etal03} 
Willott, C.~J., McLure, R.~J., \& Jarvis, M.~J.\ 2003, \apjl, 587, L15 
\bibitem[Yoshida et al.(2003)]{yos-etal03}
Yoshida, N., Sokasian, A., Hernquist, L., \& Springel, V.\ 2003, 
\apjl, 591, L1 
\bibitem[Zel'Dovich(1970)]{zel70} 
Zel'Dovich, Y.~B.\ 1970, \aap, 5, 84 

\end{thebibliography}
\end{document}